\documentclass[12pt]{article}
\usepackage{epsfig}
\usepackage{graphicx}

\def\beq{\begin{equation}}
\def\eeq#1{\label{#1}\end{equation}}
\def\eeqn{\end{equation}}

\def\beqa{\begin{eqnarray}}
\def\eeqa#1{\label{#1}\end{eqnarray}}
\def\eeqan{\end{eqnarray}}

\let\bar=\overbar

\def\L{{\cal L}}

\def\Dslash{\not{\hbox{\kern-4pt $D$}}}
\def\dslash{\not{\hbox{\kern-2pt $\del$}}}

\def\msb{{\bar{\ssstyle M \kern -1pt S}}}

\usepackage{fancyhdr,graphicx}
\def\Title#1{\begin{center} {\Large {\bf #1} } \end{center}}

\begin{document}

\Title{Bayesian analysis for two-parameter hybrid EoS with high-mass compact star twins}
% and Supernovae}

\bigskip\bigskip

%+\addcontentsline{toc}{chapter}{{\it D. Blaschke}}
%+\label{BlaschkeDavid}

\begin{raggedright}

{\it 
 D.~E.~Alvarez-Castillo$^{1,2}$, A.~Ayriyan$^{3}$, D.~Blaschke$^{1,4}$, H.~Grigorian$^{3,5}$\\
%\thanks{\tt Email: blaschke@ift.uni.wroc.pl}
\bigskip
$^{1}$Bogoliubov Laboratory for Theoretical Physics,
Joint Institute for Nuclear Research,
Joliot-Curie Str. 6,
141980 Dubna,
Russia\\
\bigskip
$^{2}$
Instituto de F\'{i}sica, 
Universidad Aut\'{o}noma de San Luis Potos\'{i},
Manuel Nava 6,
78290 SLP,
M\'{e}xico\\
\bigskip
$^{3}$Laboratory for Information Technologies,
Joint Institute for Nuclear Reseaarch,
Joliot-Curie Str. 6,
141980 Dubna,
Russia\\
\bigskip
$^{4}$Institut Fizyki Teoretycznej,
Uniwersytet Wroc{\l}awski,
pl. Maxa Borna 9,
50-204 Wroc{\l}aw,
Poland\\
\bigskip
$^{5}$Department of Physics,
Yerevan State University,
Alek Manukyan Str. 1,
Yerevan 0025,
Armenia\\
%{\tt Email: hovik.grigorian@gmail.com}}
}

\end{raggedright}

\vskip 5mm 
\centerline{\bf Abstract} 
We perform a Bayesian analysis in the basis of a recently developed two-parameter class of hybrid equations of state that allow for high-mass compact star twins.
While recently a wide range of radii, from 9 - 15 km, has been inferred for different neutron stars using different techniques, we perform our analysis under the supposition that the radii are towards the large 
end  ($13-15$ km).
We use this radius constraint together with the undebated statistically independent constraint for high masses ($\sim 2~M_\odot$) as priors in selecting the most probable hybrid equations of state from a family with two free parameters: the baryon excluded volume in the hadronic phase and the 8-quark vector channel interaction in the quark matter phase.  
\vskip 10mm

\section{Introduction}
The study of the internal composition of neutron stars is an active field of research which relies on astrophysical observations that allow to refine theoretical models. 
In this respect, the recent observations of massive neutron stars
\cite{Demorest:2010bx,Antoniadis:2013pzd} have imposed important constraints on the stiffness of the equation of state (EoS), and therefore on the density ranges
covered by the density profiles of such high-mass compact star interiors. 
On the other hand radius measurements are still far from being precise, with most of the methods relying 
on either indirect measurements or model dependent assumptions like, e.g., for the neutron star atmospheres. 
While there is a wide range of claimed radii starting from, e.g., $\sim 9$ km \cite{Guillot:2013wu}
to $\sim 15$ km \cite{Catuneanu:2013pz} we will proceed here on the assumption that the actual radii
are large, as reported by~\cite{Bogdanov:2012md,Hambaryan:2014}. 
For a recent review of astrophysical constraints on dense matter see, e.g., Ref.~\cite{Miller:2013tca}.  
 
In this contribution, we present a Bayesian analysis (BA) study case with a class of hybrid EoS characterized by two parameters.
The first one stands for the baryonic exluded volume that determines the stiffness of hadronic matter at 
high densities. 
The second one is the coupling strength for an 8-quark vector current interaction which regulates the stiffness of the high-density quark matter phase.
It turns out that within the range of variation for the excluded volume parameter there is a qualitative change in the mass-radius relation for high-mass stars: beyond a certain value for the excluded volume
the first oder phase transition proceeds with a sufficiently large jump in the energy density to cause an instability which, thanks to the stiffness of the quark matter phase at high densities goes over to a stable sequence of hybrid stars, the so-called "third family" of  compact stars.
In this situation, the conditions are fulfilled for the high-mass twin phenomenon, where stars on the high-mass end of the second family of purely hadronic neutron stars are degenerate in mass with their  twin stars on the lower mass part of the third family of hybrid stars bearing a quark matter core, see  
\cite{Benic:2014jia} and references therein. 
For a recent classification of stable hybrid star sequences under generic conditions for the EoS, see
Ref.~\cite{Alford:2013aca}.
We note that the high-mass twin phenomenon is quite substantially based on a stiffening of both, the 
hadronic and the quark matter EoS towards higher densities; it is not obtained within a systematic 
parameter scan of hadronic vs. NJL quark matter EoS \cite{Klahn:2013kga} which lacks additional stiffening effects as those introduced in Ref.~\cite{Benic:2014jia}.

This possibility of high-mass twin stars is of great importance not only due to the possibility of identification of a critical endpoint in the QCD phase diagram~\cite{Blaschke:2013ana,Alvarez-Castillo:2015xfa} but also because it provides a solution to several issues discussed in~\cite{Blaschke:2015uva}: 
the hyperon puzzle~\cite{Baldo:2003vx}, the masquerade problem~\cite{Alford:2004pf} and the reconfinement case~\cite{Lastowiecki:2011hh,Zdunik:2012dj}. 
Moreover, the transition between twin stars bears an energy reservoir \cite{Alvarez:2015}
that qualifies it as a possible engine for most energetic explosive astrophysical phenomena like gamma-ray bursts and fast radio bursts or play a role in contributing to the complex mechanism of core collapse supernova explosions. 
The present BA in the restricted two-dimensional parameter space of the new hybrid EoS
performed with the modern high mass and large radius priors will give an answer to the question whether
the high-mass twin star case is preferable over the connected hybrid star branch alternative. 

\section{Hybrid compact stars: hadronic and quark EoS}
For the neutron star description we utilize the density dependent relativistic meanfield EoS (DD2 
\cite{Typel:1999yq,Typel:2009sy}) with an excluded volume correction that takes into account the quark substructure of nucleons (protons and neutrons) leading to Pauli blocking effects. 
This correction is applied at suprasaturation densities and has the effect of stiffening the EoS without modifying any of the experimentally well constrained properties below and around saturation. 
In order to quantify the excluded volume effects, we introduce the closest packing parameter 
$\nu= 100 \times \left[n_{\nu}/{\rm fm}^{-3}\right]$, where $n_\nu$ is the closest packing density.
Increasing the excluded volume leads to a lowering of the closest packing parameter and therefore to
a lowering of the critical pressure and energy density where the first order phase transition to the high density quark matter phase occurs, implemented in the form of a Maxwell construction. 

This high-density phase is described by a NJL EoS with
multiquark interactions as introduced in~\cite{Benic:2014iaa}. 
In this model the $\eta_4$ parameter (related to the vector channels) serves to stiffen the EoS at higher densities, an important effect that
helps to support hybrid stars as massive as the 2~M$_{\odot}$ stars recently detected.

%The special selection of parameters for the hybrid stars guarantees that only massive twins will be included. 

As for the observational constraints we retain the mass and radius measurements while ignoring the baryonic mass constraint of Podsiadlowski et al. \cite{Podsiadlowski:2005ig} or Kitaura et al. 
\cite{Kitaura:2005bt} that were included in our earlier Bayesian analyses of 
Refs.~\cite{Alvarez-Castillo:2014xea,Blaschke:2014via}. 
%Although the class of EoS models discussed here does not fulfil are in
%great disagreement with Kitaura and Posiadlowski baryonic mass derivations (due to their stiffness which translates to large radii) these baryonic masses also 
%disagree with large radius values. A careful investigation redefining the methods is being in progress [Janka, personal communication]. This is the reason why in this
%seminal, two dimensional Bayesian analysis we have decided to put baryonic masses aside.

In the BA by Steiner et al. \cite{Steiner:2010fz} the luminosity radius 
extracted for burst sources has been used to  constrain a combined mass-radius 
relationship. 
This method is problematic, in particular because of the unknown stellar 
atmosphere composition, uncertainties in the distance to the 
source, the bias of the parabolic M-R constraint with the shape of stellar
sequences in the M-R diagram for typical EoS and last but not least due to unknown 
details of the burst mechanism. 

\section{Bayesian analysis technique}

We start by defining a vector of free parameters 
$\overrightarrow{\pi}=\{\nu,\eta_4\}$, 
%where $\varepsilon_H$ is the critical value of energy density at the onset of 
%the phase transition (PT), $\gamma = \epsilon / \epsilon_c^{I}$ is a ratio of 
%the energy jump on PhT to the critical one, and ${c}_{s}^{2}$ is square of 
%speed of sound in quark matter. These parameters are 
which correspond to all the possible models with phase transition from nuclear to quark matter using the EoS described above. 
The way we sample these parameters is
\begin{equation}
\label{pi_vec}
\overrightarrow{\pi}_i = \left\{\nu_{(k)},\eta_{4(l)}\right\},
\end{equation}
where $i = 0\dots N-1$ with $N = N_1\times N_2$ 
such that $i = N_2\times k + l$ and $k = 0\dots N_1-1$, $l = 0\dots N_2-1$, with $N_1$ and $N_2$ 
being the total number of parameters $\nu_{(k)}$ and $\eta_{4 (l)}$, respectively.

Once all the models are defined the neutron star configurations are obtained by solving the 
Tolman-Oppenheimer-Volkoff (TOV)
equations \cite{Tolman:1939jz,Oppenheimer:1939ne}. 
These results allow us to use different neutron star observations in order 
to determine the probability that a given EoS model fulfils the observational constraints. 
We use here a mass constraint~\cite{Antoniadis:2013pzd} and 
a radius constraint~\cite{Bogdanov:2012md}.

Our goal is to find the set of most probable $\overrightarrow{\pi}_i$ matching the above
constraints using the BA technique.
For initializing the BA we propose that {\it a priori} each vector of 
parameters $\overrightarrow{\pi}_i$ has the same probability, $P\left(\overrightarrow{\pi}_i\right) = 1/N$, 
for all $i$.

\subsection{Mass constraint} 
We describe the error for the event $E_A$ of a mass measurement of the high-mass pulsar 
PSR~J0348+0432~\cite{Antoniadis:2013pzd} with a normal distribution 
$\mathcal{N}(\mu_A,\sigma_A^2)$, where the mean value of the mass is 
$\mu_A = 2.01~\mathrm{M_{\odot}}$ and the variance is
$\sigma_A = 0.04~\mathrm{M_{\odot}}$. 
Using this assumption we compute the conditional probability of the 
event $E_{A}$ (under the condition that the neutron star is described by the EoS model with 
the parameters $\overrightarrow{\pi}_i$) with
\begin{equation}
\label{p_anton}
P\left(E_{A}\left| \overrightarrow{\pi}_i\right.\right) = \Phi(M_i, \mu_A, \sigma_A)~.
\end{equation}
Here $M_i$ is the maximum mass accessible with the vector $\overrightarrow{\pi}_i$ and 
$\Phi(x, \mu, \sigma)$  is the cumulative distribution function for the Gaussian distribution
\begin{equation}
\label{Laplas}
\Phi(x, \mu, \sigma) = 
\frac{1}{2}
\left[1+{\rm erf}\left(\frac{x-\mu}{\sqrt{2\sigma^2}}\right)\right].
\end{equation}

\subsection{Radius constraint}    
We consider here a very promising technique to measure radii of neutron stars that is based on the 
pulse phase resolved X-ray spectroscopy which properly accounts for the system geometry of a radio pulsar. 
This radius measurement gives $\mu_B = 15.5~\mathrm{km}$ and 
$\sigma_B = 1.5~\mathrm{km}$ for PSR~J0437-4715~\cite{Bogdanov:2012md}. 
Moreover, Hambaryan et al. \cite{Hambaryan:2014} have also reported compatible radius measurements 
for RXJ 1856.5-3754.
 
We compute the conditional probability of the event 
$E_{B}$ that the measured radius of the neutron star corresponds to the model with 
$\overrightarrow{\pi}_i$ as
\begin{equation}
\label{p_bogdan}
P\left(E_{B}\left|\overrightarrow{\pi}_i\right.\right) = \Phi(R_i, \mu_B, \sigma_B)~.
\end{equation}
Here the value $R_i$ is the maximum radius for the given vector $\overrightarrow{\pi}_i$.

\subsection{Calculation of {\it a posteriori} probabilities}
It is important to note that these measurements are independent of each other. 
This means that we can compute the complete conditional probability of an 
event $E$ given $\overrightarrow{\pi}_i$ that corresponds to the product of the conditional 
probabilities of all measurements, in our case resulting from the 
constraints $E_A$, $E_B$, 
\begin{equation}
\label{p_event}
P\left(E\left|\overrightarrow{\pi}_i\right.\right) = 
P\left(E_{A}\left|\overrightarrow{\pi}_i\right.\right) 
\cdot P\left(E_{B}\left|\overrightarrow{\pi}_i\right.\right).
\end{equation}
Thus, we can derive the probability of the measurement of an EoS represented by a vector of parameters
$\overrightarrow{\pi}_i$ using Bayes' theorem
\begin{equation}
\label{pi_apost}
P\left(\overrightarrow{\pi}_i\left|E\right.\right) = 
\frac{P\left(E\left|\overrightarrow{\pi}_i\right.\right)
P\left(\overrightarrow{\pi}_i\right)}{\sum\limits_{j=0}^{N-1}P\left(E\left|\overrightarrow{\pi}_j\right.\right)P\left(\overrightarrow{\pi}_j\right)}.
\end{equation}
Since all $P(\overrightarrow{\pi}_i)=1/N$ and $P(\overrightarrow{\pi}_i|E)$ is normalized to
$\sum_j P(E|\overrightarrow{\pi}_j)=\widetilde{N}$, we have 
\begin{equation}
P(\overrightarrow{\pi}_i|E)=P(E|\overrightarrow{\pi}_i)/\widetilde{N}~.
\end{equation}
\begin{figure}[htb]
\centering
\includegraphics[width=0.8\textwidth]{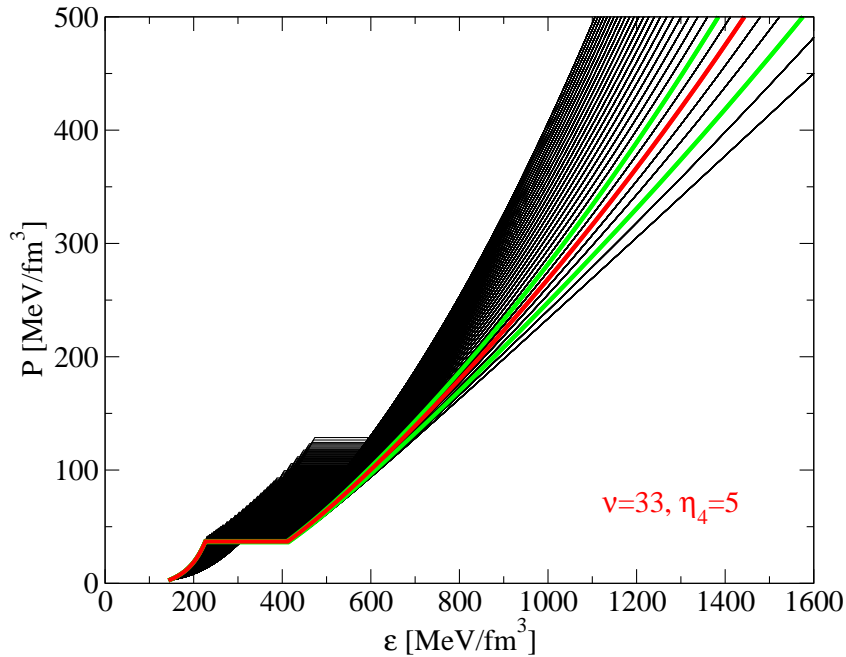}
\caption{Family of hybrid EoS (black lines) with a first order phase transition between hadronic matter described by the excluded volume corrected DD2 relativistic meanfield EoS and quark matter described by a NJL model with higher order (8-quark) interactions that provide the important stiffening at high densities due to a large coupling strength $\eta_4$ in the vector current channel, see \cite{Benic:2014jia}. 
The optimal EoS (bold red line) corresponds to an available volume parameter $\nu=33$ and $\eta_4=5$.
The limits of the $1\sigma$ range are indicated by bold green lines.}
\label{fig:p-e}
\end{figure}

\section{Results and discussion}

In order to cover a relevant set of possible hybrid compact star EoS in the pressure-energy density plane
we have varied the closest packing parameter in the range $\nu= 33, \dots, 100$ corresponding to closest packing densities of $ n_{\nu} = 0.33, 0.34,\dots , 1.00$ fm$^{-3}$,
and the 8-quark vector current coupling in the range $\eta_4=0,1,2,\dots, 30$. 

The hybrid EoS resulting from Maxwell constructions between all combinations of hadronic and quark
matter EoS in this two-dimensional parameter space is shown in Fig.~\ref{fig:p-e}.

\begin{figure}[htb]
\centering
\includegraphics[width=\textwidth]{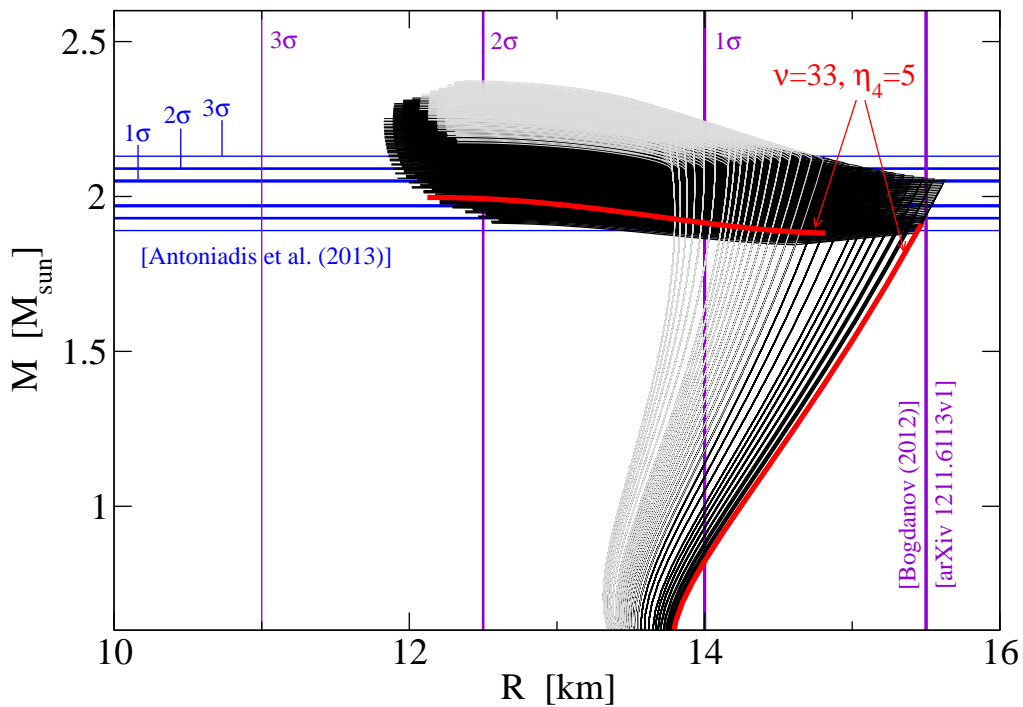}
\caption{Hybrid compact star sequences described by the family of hybrid EoS in Fig.~\ref{fig:p-e} used for Bayesian studies together with observational high precision mass 
(1$\sigma$, 2$\sigma$ and 3$\sigma$ horizontal bands) \cite{Antoniadis:2013pzd} and radius measurements (1$\sigma$, 2$\sigma$ and 3$\sigma$ vertical bands)~\cite{Bogdanov:2012md}.}
\label{fig:MvsR}
\end{figure}

Fig.~\ref{fig:MvsR} shows the resulting compact star sequences in the mass-radius plane which have a vertical branch corresponding to hadronic stars and an almost horizontal branch for hybrid stars with a
quark matter core. 
The parameters are chosen in such a range that their variation entails that the hybrid star branch 
becomes disconnected from the hadronic one due to the appearance of a set of unstable configurations.
This characterizes the appearance of high-mass twin stars and happens in particular when varying the closest packing parameter $\nu$ (i.e., the excluded volume)  in all of these models.

We have already observed in \cite{Benic:2014jia} that increasing the $\eta_4$ parameter for a fixed $\nu$ value corresponds to an increase of the maximum mass, while the difference between the radius of the hybrid star and its hadronic twin decreases. 
This general behaviour is preserved for any fixed $\nu$.

\begin{figure}[htb]
\centering
\includegraphics[width=0.6\textwidth]{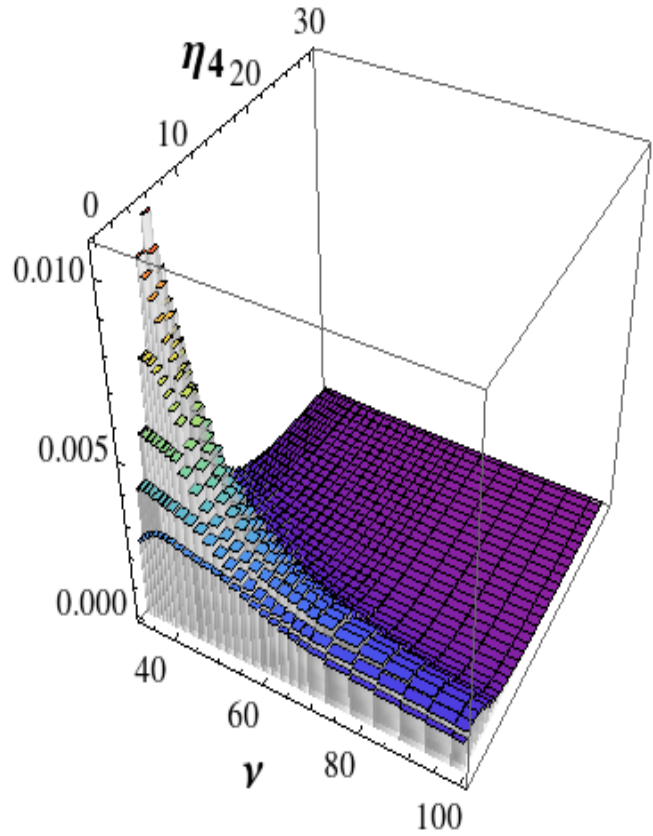}
\caption{A posteriori probabilities derived from a two-dimensional Bayesian analysis. 
The vertical axis shows the probability for an EoS characterized by the parameters
$\nu$ and $\eta_4$ to fulfill the observational constraints for mass and radius.}
\label{fig:LegoPlots}
\end{figure}

{\it A posteriori} probabilities of identification for these twin EoS are shown in Fig.~\ref{fig:LegoPlots}. 
The models with higher probabilities correspond to the ones that predict
large radii and are associated to small closest packing parameters $\nu$.
The optimal value of the vector coupling strength is $\eta_4=5$.
We note that the family of hybrid EoS accessible within the given two-dimensional parameter space 
describes sequences of compact stars that have a hadronic branch and a hybrid star branch.
The latter appears connected to the former for not too large excluded volumina, i.e. for larger closest packing parameter $\nu$.
For sufficiently small closest packing parameters the compact stars on the hadronic branch have radii 
exceeding $14$ km and the phase transition proceeds with a sufficiently large jump in the energy density 
so that the hybrid star sequence gets disconnected from the hadronic one, forming a so-called "third family".    
We summarize our conclusions from this study:
\begin{enumerate}
\item The most probable models exhibit high-mass twin star configurations with quite distinguishable radii,
differing by about 2 km.
\item The region of the most probable models in the two-dimensional parameter space is sufficiently narrow, covering the ranges $33<\nu < 38$ and $3<\eta_4< 7$.
\item The most probable models have a relatively small closest packing parameter $\nu\sim 33$ and a 
not too large vector coupling strength $\eta_4\sim 5$.
\item The existence of the horizontal branch signals a strong first order deconfinement phase transition and is a feature accessible to verification by observation. To that end, at least for two high-mass pulsars with masses $\sim 2~M_\odot$ (like PSR J1614-2230 and PSR J0348+0432) the radii should be measured to sufficient accuracy and turn out to be significantly different.  
\end{enumerate}
The next two steps in the development of the approach are devoted to an improvement of the variability of the dense matter EoS within a two-dimensional parameter space embodying, e.g., also the purely hadronic case without a phase transition and to mimicking the occurrence of structures (so-called "pasta phases") in the phase transition region \cite{Yasutake:2014oxa,Alvarez-Castillo:2014dva}. 

\subsection*{Acknowledgements}

We express our thanks to the organizers of the CSQCD IV conference for providing an 
excellent atmosphere which was the basis for inspiring discussions with all participants. 
We thank Stefan Typel and Sanjin Benic for providing the necessary EoS data sets for this work
and for their comments to this manuscript.
We have greatly benefitted from discussions with Cole Miller on the radius constraints. 
This work was supported  by the Polish NCN under grant No. UMO-2014/13/B/ST9/02621.
D.E.A-C. and H.G. acknowledge support by the programme for exchange between JINR Dubna and 
Polish Institutes (Bogoliubov-Infeld programme) by the COST Action MP1304 "NewCompStar"
D.B. was supported in part by the Hessian LOEWE initiative through NIC for FAIR.

\end{document}